\documentclass{article}
\usepackage{spconf,amsmath,graphicx}
\usepackage{caption}
\usepackage{setspace}
\usepackage{epstopdf}
\usepackage[]{algorithm2e}
\usepackage{amsmath}

\title{Using a Power Law Distribution to describe Big Data}
\name{Vijay Gadepally, Jeremy Kepner\thanks{This work is sponsored by the
  Assistant Secretary of Defense for Research and Engineering under
  Air Force Contract \#FA8721-05-C-0002.  Opinions, interpretations,
  recommendations and conclusions are those of the authors and are not
  necessarily endorsed by the United States Government}}
\address{MIT Lincoln Laboratory, Lexington, MA 02420 \\ 
\{vijayg, kepner\}@ ll.mit.edu}

\begin{document}
%
\maketitle
%


\begin{abstract}
The gap between data production and user ability to access, compute and produce
meaningful results calls for tools that address the
challenges associated with big data volume, velocity and variety.  One of the key
hurdles is the inability to methodically remove expected or
uninteresting elements from large
data sets. This difficulty often
wastes valuable researcher and computational time by expending resources on
uninteresting parts of data. Social sensors, or sensors which produce
data based on human activity,
such as Wikipedia, Twitter, and Facebook have an underlying
structure which can be thought of as having a Power Law
distribution. Such a distribution implies that few nodes generate
large amounts of data. In this article, we propose a technique to take an
arbitrary dataset and compute a power law distributed background
model that bases its parameters on observed statistics. This model
can be used to determine the suitability of using a power law
or automatically identify high degree nodes for filtering and can be
scaled to work with big data.
\end{abstract}
\section{Introduction}
\label{sec:intro}

The 3 V's of big data: volume, velocity
and variety~\cite{laney20013d}, provide a guide to the outstanding challenges associated with
working with big data systems. Big data \textbf{volume} stresses the storage,
memory and compute capacity of a computing system and requires access
to a computing cloud. The \textbf{velocity} of big data stresses the rate at which data can
be absorbed and meaningful answers produced. Big data \textbf{variety}
makes it difficult to develop algorithms and tools that can address
the large diversity  of input data. One of the key challenges is in
developing algorithms that can provide analysts with basic knowledge
about their dataset when they have little to no knowledge of the data
itself.

In~\cite{gadepally2014big}, we proposed a series of steps that an
analyst should take in analyzing an unknown dataset including a technique
similar to spectral analysis - Dimensional Data Analysis (DDA). As a next step in the data analysis
pipeline, we propose determining a suitable background model. Big data
sets come from a variety of sources such as social media, health care
records, and deployed sensors, and the background model for such datasets
are as varied as the data itself. One of the popular statistical
distributions used to explain such data is the power law
distribution. There is much related work in the area. Studies such as
~\cite{newman2005power,gabaix1999zipf,adamic2002zipf,nyt} have
looked at power laws as underlying models for
human generated big data collected from sources such as social media,
network activity and census
data. However, there has also been some controversy that
while data may look like it follows a power law it may in fact be
better described by other distributions such as exponential or log-normal
distributions~\cite{mitzenmacher2004brief,blog}. While a power law distribution may seem a fitting background model for
an observed dataset, large fluctuations in lower degree terms (the
tail of the distribution) may skew the estimation of power law
parameters~\cite{clauset2009power}. Further, the estimation of power
law exponent can be heavily dependent on decisions such as
binning~\cite{white2008estimating} which may lead to problems such as estimator bias~\cite{goldstein2004problems}. 

We propose a
technique that takes an unknown dataset, estimates the parameters and
binning of the degree distribution of an power law distributed dataset that follows 
constraints enforced by the observed dataset, and aligns the degree
distribution of the observed
dataset to the structure of the perfect power law distribution in order to provide a
clear view into the applicability of a power law model. 

The article is structured as follows: Section~\ref{sec:theory} describes
the relationship between signal processing and big data;
Section~\ref{sec:technique} describes the proposed technique to
estimate the power law parameters of a dataset; Section~\ref{sec:app}
describes application examples. Finally, we conclude in Section~\ref{sec:conc}.

\section{Signal Processing and Big Data}
\label{sec:theory}

Detection theory in signal processing is the ability to discern
between different signals based on the statistical properties of the
signals. In the simplest form, the decision is to choose whether a
signal is present or absent. In making this determination,
the observed signal is compared against the expected background signal
when no signal is present. A deviation from this background model indicates
the presence of a signal. While it is common to represent the
background model as additive white gaussian noise (AWGN), the 
model may change depending on physical factors such as
channel parameters or known noise characteristics.

Big data can be as a considered a high dimensional signal that is
projected to an n-dimensional space. Big data, similar to the 1-, 2-
and 3-D signals that signal processing has traditionally dealt with, are
often noisy, corrupted, or misaligned. The concept of \textit{noise} in big
data can be thought of as unwanted information which impairs the detection of
activity or important entities present in the dataset. For example, consider a situation in
which network logs are collected to determine
foreign or dangerous connections out of a
network. Detecting such activity may be difficult due to the presence
of few
vertices (such as connections to www.google.com) with a very large number of
connections (edges). This information can be considered the equivalent of stop words
in text
analytics~\cite{zhao2011comparing,kouloumpis2011twitter}. While such
data may help form a useful statistic, often, these
entries impair the ability to find activity of interest that occurs at
a lower activity threshold. Often,
such vertices with a large number of edges (high degree vertices) are manually removed based on empirical
evidence in order to improve the big data \textit{signal to noise
  ratio}. Knowledge of a
suitable big data background model can highlight such vertices and help
with the automated removal of components in the dataset which are of
minimal interest. This concept parallels the concept of filtering in signal
processing. In fact, such parallels between big data and signal
processing kernels are numerous. In~\cite{sandryhaila2014big}, for example, the authors look
at the commonality between certain signal processing kernels and
graph operations. As another example, the authors of
~\cite{Muchnik:2013aa} study data collected from social networks
and their underlying statistical distribution.

\begin{figure*}
\centerline{
\includegraphics[width=7.1in]{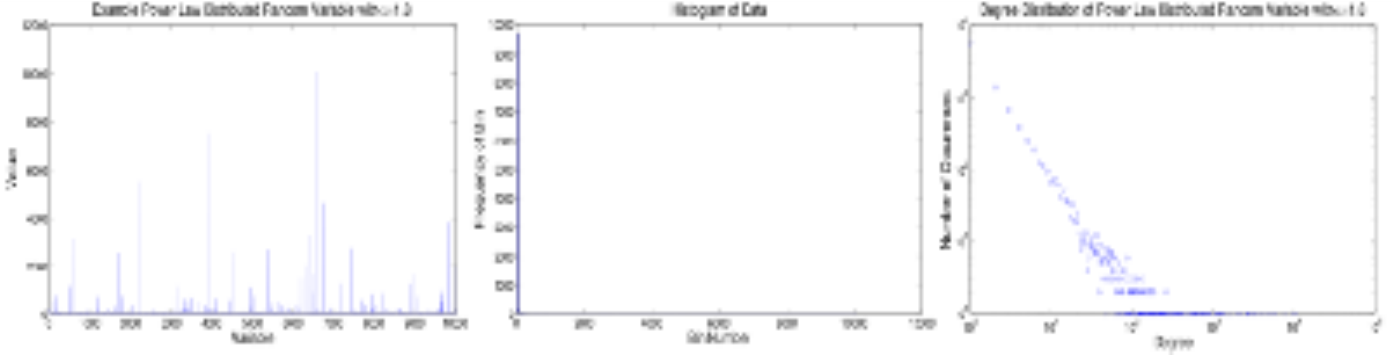}
}
\caption{Computing the histogram and degree distribution of 10,000
  data points in which the magnitude was determine by drawing from a
  power law distribution with $\alpha=1.8$.}
\label{raw2degree}
\end{figure*}

A common distribution that is observed in many datasets is the power
law distribution. A power law distribution is one in which a small
number of vertices are incident to a large number of edges. This
principle has also been referred to as the Pareto principle, Zepf's
Law, or the 80-20 rule. A power law
distribution for a random variable \textbf{x}, is defined to be:

\begin{eqnarray}
p(x)=cx^{-\alpha} \nonumber \end{eqnarray}

\noindent where the exponent $\alpha$ represents with the power of the
distribution. An illustrative example of 10,000 randomly generated
points drawn from a power law distribution with exponent $\alpha=1.8$
is shown in Figure~\ref{raw2degree}. The middle figure shows the
histogram of such a random variable. Finally, the rightmost
image shows the degree distribution of the signal. Note the large number
of low degree nodes and small number of high degree nodes. While a power law distribution may be applicable to a given dataset
based on physical characteristics or empirical evidence, a more
rigorous approach is needed to fit observed data to a power law model
in order to verify the applicability of a power law distribution.

\section{Power Law Modeling Technique}
\label{sec:technique}

This section describes the proposed technique for comparing an
observed dataset with an ideal power law distributed dataset whose
parameters are derived from a statistical analysis of the observed dataset.

\subsection{Definitions}
\label{definitions}

A large dataset can be represented by a graph through the
adjacency matrix representation or incidence matrix
representation~\cite{fulkerson1965incidence}. An adjacency matrix has dimensions
$n \times n$, where $n$ corresponds to the number of vertices in the graph. A vertex out-degree is a
count of the number of edges in a directed graph which leave a
particular vertex. The vertex in-degree, on the contrary, is a count
of the number of edges in a directed graph which enter a particular
vertex. A popular way to represent the in-degree and
out-degree distributions is through the degree distribution which is a
statistic that computes the number of vertices that are of a certain
degree. Such a count is very relevant to techniques such as graph
algorithms and social networks.

\begin{figure}[h!]
\centerline{
\includegraphics[width=3.45in]{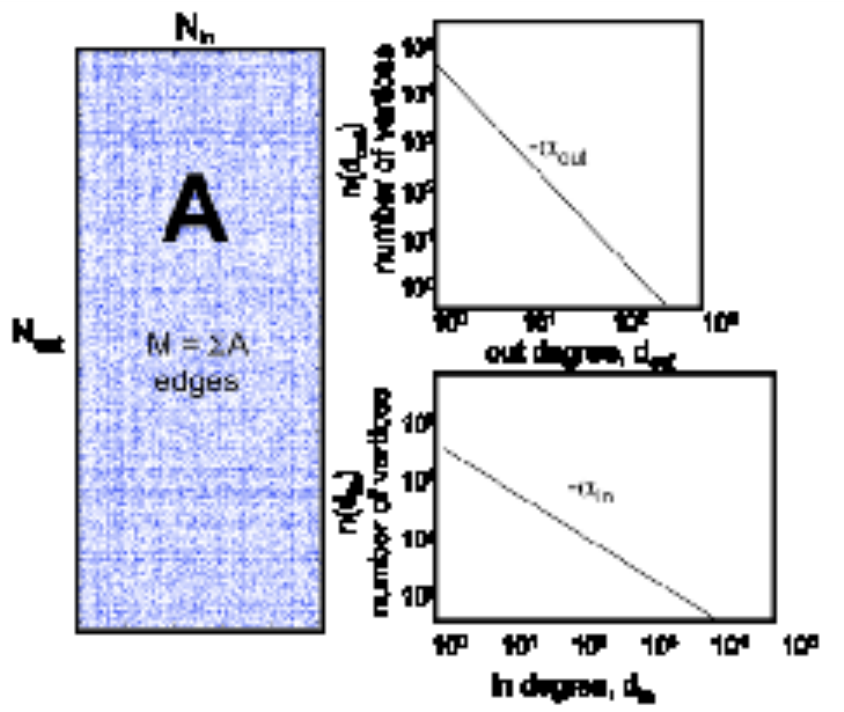}
}
\caption{Adjacency matrix for notional power law distributed random graph with
  corresponding in and out degree distributions.}
\label{inandoutdegree}
\end{figure}

The in-degree and out-degree of a graph can be determined by the
following relations:

\begin{eqnarray}
d_{in}(i) = \sum_{j}E_{ij} \nonumber \\
d_{out}(i) = \sum_{j} E_{ji} \nonumber
\end{eqnarray}

\noindent where $d_{in}(i)$ and $d_{out}(i)$ represent the in and out degree
at bin $i$ and $E$ represents the graph incidence matrix. The selection of the the total number of bins ($N_{d}$) is an
important factor in determining the degree distribution given by
$d_{in}$, $d_{out}$, and $i \leq
N_{d}$. An illustrative example is shown in
Figure~\ref{inandoutdegree} to demonstrate the parameters described in
this section.

The degree distribution conveys many important
pieces of information. For example, it may show that a large number of vertices have a
degree of 1, which implies that a majority of vertices are connected
only to one other vertex. Further, it may show that there are a small
number of vertices with high
degree (many edges). In a social media dataset, such vertices may correspond to
popular users or items.

The maximum degree vertex is said to be $d_{N_{d}}=d_{max}$. The total number of vertices ($N$) and
edges ($M$) can be computed as:

\begin{eqnarray}
N = \sum_{i=1}^{N_{d}} n(d_{i}) \nonumber \\
M = \sum_{i}^{N_{d}} n(d_{i}) * d_{i} \nonumber \end{eqnarray}

\noindent where $n(d_{i})$ is defined as number (count) of vertices with degree $d_{i}$.

\subsection{Power Law Fitting}
\label{powerlawfitting}

\medskip
\noindent \textbf{Step 1: Find parameters of observed data}
\smallskip

In order to determine the power law parameters for an arbitrary data set, data is
first converted to an adjacency matrix representation. In the
adjacency matrix, row and columns represent vertices with outbound
and inbound edges respectively.  A non-zero entry at a particular
row and column pair indicates the existence of an edge between two
vertices. Often, data may be collected and stored in an incidence
matrix where rows of the matrix represent edges and columns represent
vertices. For data in an intermediate format such as
the incidence matrix ($E$), it is possible to convert this
representation to the adjacency
matrix ($A$) using the following relation:

\begin{eqnarray}
A = |E < 0|^{T} * |E > 0| \nonumber
\end{eqnarray}

\noindent where $|E<0|$ and $|E>0|$ represent
the incidence matrix with outbound and inbound edges only. From the
adjacency matrix, it is possible to calculate
the degree distribution as described to extract the
parameters $\alpha$, the vertex with maximum degree $d_{max}$, the number of
vertices with exactly 1 edge, $n(d_{1})$, and the number of bins
$N_{d}$. There are many proposed methods to calculate the power law exponent ~\cite{bauke2007parameter,
  white2008estimating}. For the purpose of this study, a simple first
order estimate of $\alpha$ is sufficient and it should satisfy the
intuitive property that the count $n(d_{1})$ and degree $d_{max}$ be
included since most natural datasets will have at least one vertex with
$degree=1$ and a vertex with large degree. Furthermore, the exponent
should take into account. Therefore, we propose the
following simple relationship to calculate $\alpha$:

\begin{eqnarray}
\label{eqn1}
\alpha = log(n(d_{1}))/log(d_{max})
\end{eqnarray}

\noindent Using the power law exponent calculated in Equation~\ref{eqn1} allows
an initial comparison between the observed dataset and a power law
distribution.

\medskip
\noindent \textbf{Step 2: Calculate ``Perfect'' Power Law Parameters}
\smallskip

Given the degree distribution of the observed data we can compute the parameters:
$\alpha^{obs}$, $d_{i}^{obs}$, $n(d_{i}^{obs})$, $N^{obs}_{d}$,
$M^{obs}$, and $N^{obs}$, using the relations provided in the
definitions section (Section~\ref{definitions}). In order to see if the observed data fits a power
law distribution, we need to be able to determine what an ideal power
law distributed dataset would look like for parameters similar to
those observed. This ideal distribution is referred to as a
``perfect'' power law distribution.

The ``perfect'' power law distribution can be determined by computing
the parameters $d_{i}$, $n(d_{i})$, and $N_{d}$ which closely
fit the observed data while also maintaining the number of vertices and edges. While theoretically, any distribution which satisfies the properties
$\alpha>0$, $d_{max}>1$ and $N_{d}>1$ can be used to form a power
law model, we also desire values which maintain the total number of
vertices and edges, $N$ and $M$. Essentially, given an observed number of vertices and
edges, compute the quantities  $N_{d}$, $n(d_{i})$ and $d_{i}$ where
$i \leq N_{d}$ that form a power law distribution (with $\alpha$)
which also satisfy the property that $M \approx M^{obs}$ and $
N \approx N^{obs}$.

These values can be solved by using a combination of optimization techniques such as
exhaustive search, simulated annealing, or Broyden's algorithm, to find the values $d_{i}$ and $n(d_{i})$ that minimize:

\begin{multline}
\min_{d,n}f(d_{i},n(d_{i})) \nonumber \\
= \sqrt{ \arrowvert M^{obs}-\sum_{i} n(d_{i}) \arrowvert^{2}
  +  \arrowvert N^{obs} - \sum_{i} n(d_{i}) * d_{i}\arrowvert^{2}} \nonumber
\end{multline}

\noindent where $M^{obs}$ and $N^{obs}$ are the observed number of edges and
vertices. From the estimate of $d_{i}$ and $n(d_{i})$
we can determine $N_{d}$ (given by
the number of output $d_{i}$), and $d_{max}$ (given by $d_{N_{d}}$).

\medskip
\noindent \textbf{Step 3:  Align observed data with background
  model}
\smallskip

The values of $N_{d}$, $d_{i}$, and $n(d_{i})$ from the previous
step, provides a power law distributed dataset with power
$\alpha$. However, the degree binning may be different from the
observed distribution. In order to compare the observed data with the background
model, it is necessary to rebin the observed data such that it aligns
with the background model. Using the rebinned observed data
(represented by the parameters $d_{i}^{rebin}$ and $n(d_{i}^{rebin})$)
it is possible to determine the power law nature of the observed
dataset. Both datasets use the same degree binning using
algorithm 1.

\begin{algorithm}
\KwData{$d_{i}$, $d_{i}^{obs}$}
\KwResult{$d_{i}^{rebin}$, $n(d_{i}^{rebin})$}
\For{i=1:$N_{d}$} {$d_{i}^{rebin}=d_{i}$ \\$n(d_{i}^{rebin})$ = number
of vertices binned to $i$}
\label{algo1}
\caption{Algorithm to rebin observed data into fitted data bins}
\end{algorithm}

\section{Application Example}
\label{sec:app}

To demonstrate the application of steps provided in Section
~\ref{powerlawfitting}, we describe two examples - a Twitter
dataset and a corpus of news articles provided by Reuters. We use the open source Dynamic
Distributed Dimensional Data Model (D4M, d4m.mit.edu) to store and
access the required data. As a first step, data is converted to the D4M schema
~\cite{kepner2013d4m}, which organizes data into an
associative array, representing data as an incidence matrix. The
Twitter dataset contains all the metadata associated with
approximately 2 million tweets. For the purpose of this example, we
have considered only a subset of the data which corresponds to Twitter
usernames. 

\begin{figure}[t!]
\centerline{
\includegraphics[width=4.8in]{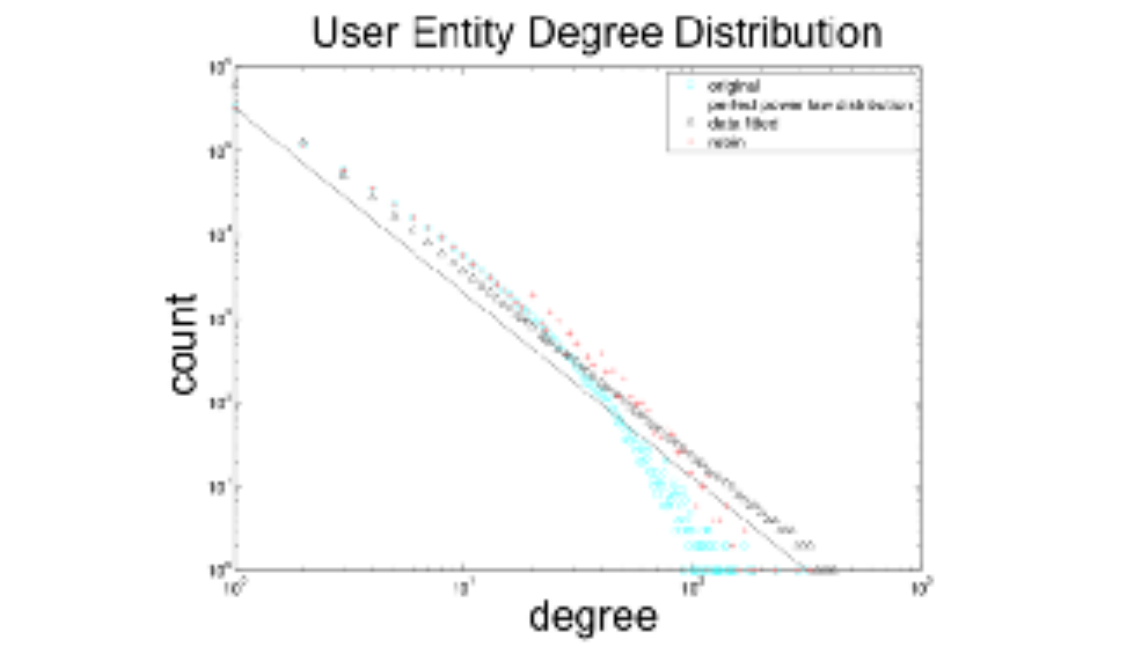}
}
\caption{Fitting a Power Law distribution to Twitter user data.}
\label{powerlawresults}
\end{figure}

To begin, we determine the adjacency matrix of the associative array
data using the relation outlined in the previous section. With the
adjacency matrix, we can determine the degree distribution of the
observed dataset as demonstrated by the blue circles in
Figure~\ref{powerlawresults}. Using the values of N and M from this
distribution, we can find the values of $d_{i}$ and $n(d_{i})$ that
fit the N and M values of the observed dataset. The obtained values
are plotted in Figure~\ref{powerlawresults} as the black
triangles. As a final step, we rebin the original degree
distribution to align with the bins of the ideal distribution. The
results of rebinning the observed distribution are shown in
Figure~\ref{powerlawresults} as red plus signs.

\begin{figure}[t!]
\centerline{
\includegraphics[width=3.45in]{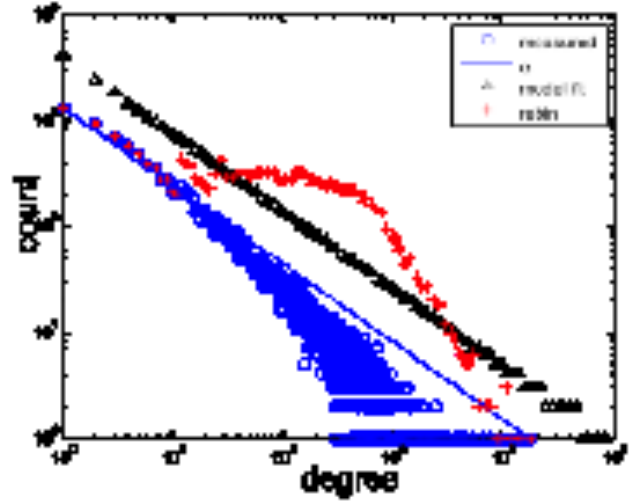}
}
\caption{Fitting a Power Law distribution to Reuters data corpus.}
\label{reuters}
\end{figure}

For the Twitter user data of Figure~\ref{powerlawresults},  we see that a power law distribution
provides a good representation of the data. The second dataset, a corpus of news articles from Reuters, seems
to follow a power law distribution. However, once we fit the perfect
power law distribution and rebin the original data, we see that the
dataset does not follow a power law distributions evidenced by the
large bulge in Figure~\ref{reuters}.

\section{Conclusions and Future Work}
\label{sec:conc}

In this article, we presented a technique to uncover the underlying
distribution of a big dataset. One of the most common statistical
distributions attributed to a variety of human generated big data
sources, such as social media, is the power law distribution. Often,
however, data that seems to adhere to a power law distribution may not
be well described by
such a distribution. In such situations, it is important to be aware
of the underlying background model of the dataset before further processing. Our future work includes investigating the big
data equivalents of sampling and big data filtering.

\section{Acknowledgements}

The authors wish to thank the LLGrid team at MIT Lincoln Laboratory
for their assisstance in setting up the experiments.



\begin{spacing}{0.95}
\bibliographystyle{IEEEbib}
\bibliography{icassp}

\begin{thebibliography}{10}

\bibitem{laney20013d}
Doug Laney,
\newblock ``3{D} data management: Controlling data volume, velocity and
  variety,''
\newblock {\em META Group Research Note}, vol. 6, 2001.

\bibitem{gadepally2014big}
Vijay Gadepally and Jeremy Kepner,
\newblock ``Big data dimensional analysis,''
\newblock {\em IEEE High Performance Extreme Computing (HPEC)}, 2014.

\bibitem{newman2005power}
Mark Newman,
\newblock ``Power laws, pareto distributions and zipf's law,''
\newblock {\em Contemporary Physics}, vol. 46, no. 5, pp. 323--351, 2005.

\bibitem{gabaix1999zipf}
Xavier Gabaix,
\newblock ``Zipf's law for cities: an explanation,''
\newblock {\em Quarterly Journal of Economics}, pp. 739--767, 1999.

\bibitem{adamic2002zipf}
Lada Adamic and Bernardo Huberman,
\newblock ``Zipf's law and the internet,''
\newblock {\em Glottometrics}, vol. 3, no. 1, pp. 143--150, 2002.

\bibitem{nyt}
Paul Krugman,
\newblock ``The power law of twitter,''
\newblock {\em
  http://krugman.blogs.nytimes.com/2012/02/08/the-power-law-of-twitter/?\_php=true\&\_type=blogs\&\_r=0},
  2014.

\bibitem{mitzenmacher2004brief}
Michael Mitzenmacher,
\newblock ``A brief history of generative models for power law and lognormal
  distributions,''
\newblock {\em Internet mathematics}, vol. 1, no. 2, pp. 226--251, 2004.

\bibitem{blog}
``Twitter followers do not obey a power law, or paul krugman is wrong,''
\newblock {\em
  http://blog.luminoso.com/2012/02/09/twitter-followers-do-not-obey-a-power-law-or-paul-krugman-is-wrong/}.

\bibitem{clauset2009power}
Aaron Clauset, Cosma~Rohilla Shalizi, and Mark Newman,
\newblock ``Power-law distributions in empirical data,''
\newblock {\em SIAM review}, vol. 51, no. 4, pp. 661--703, 2009.

\bibitem{white2008estimating}
Ethan White, Brian Enquist, and Jessica Green,
\newblock ``On estimating the exponent of power-law frequency distributions,''
\newblock {\em Ecology}, vol. 89, no. 4, pp. 905--912, 2008.

\bibitem{goldstein2004problems}
Michel~L. Goldstein, Steven~A. Morris, and Gary~G. Yen,
\newblock ``Problems with fitting to the power-law distribution,''
\newblock {\em The European Physical Journal B-Condensed Matter and Complex
  Systems}, vol. 41, no. 2, pp. 255--258, 2004.

\bibitem{zhao2011comparing}
Wayne~Xin Zhao, Jing Jiang, Jianshu Weng, Jing He, Ee-Peng Lim, Hongfei Yan,
  and Xiaoming Li,
\newblock ``Comparing twitter and traditional media using topic models,''
\newblock in {\em Advances in Information Retrieval}, pp. 338--349. Springer,
  2011.

\bibitem{kouloumpis2011twitter}
Efthymios Kouloumpis, Theresa Wilson, and Johanna Moore,
\newblock ``Twitter sentiment analysis: The good the bad and the omg!,''
\newblock {\em Icwsm}, vol. 11, pp. 538--541, 2011.

\bibitem{sandryhaila2014big}
Aliaksei Sandryhaila and Jose Moura,
\newblock ``Big data analysis with signal processing on graphs: Representation
  and processing of massive data sets with irregular structure,''
\newblock {\em Signal Processing Magazine, IEEE}, vol. 31, no. 5, pp. 80--90,
  2014.

\bibitem{Muchnik:2013aa}
Lev Muchnik, Sen Pei, Lucas~C. Parra, Saulo D.~S. Reis, Jos{\'e}. Andrade~Jr,
  Shlomo Havlin, and Hern{\'a}n~A. Makse,
\newblock ``Origins of power-law degree distribution in the heterogeneity of
  human activity in social networks,''
\newblock {\em Nature Scientific Reports}, vol. 3, 05 2013.

\bibitem{fulkerson1965incidence}
Delbert Fulkerson and Oliver Gross,
\newblock ``Incidence matrices and interval graphs,''
\newblock {\em Pacific journal of mathematics}, vol. 15, no. 3, pp. 835--855,
  1965.

\bibitem{bauke2007parameter}
Heiko Bauke,
\newblock ``Parameter estimation for power-law distributions by maximum
  likelihood methods,''
\newblock {\em The European Physical Journal B-Condensed Matter and Complex
  Systems}, vol. 58, no. 2, pp. 167--173, 2007.

\bibitem{kepner2013d4m}
Jeremy Kepner, Christian Anderson, William Arcand, David Bestor, Bill Bergeron,
  Chansup Byun, Matthew Hubbell, Peter Michaleas, Julie Mullen, David O'Gwynn,
  et~al.,
\newblock ``{D4M} 2.0 schema: A general purpose high performance schema for the
  accumulo database,''
\newblock in {\em IEEE High Performance Extreme Computing Conference (HPEC)},
  2013.

\end{thebibliography}
\end{spacing}


\end{document}